# Atomic and electronic reconstruction at van der Waals interface in twisted bilayer graphene

Hyobin Yoo[1], Rebecca Engelke[1], Stephen Carr[1], Shiang Fang[1], Kuan Zhang[2], Paul Cazeaux[3], Suk Hyun Sung[4], Robert Hovden[4], Adam W. Tsen[5], Takashi Taniguchi[6], Kenji Watanabe[6], Gyu-Chul Yi[7], Miyoung Kim[8], Mitchell Luskin[9], Ellad B. Tadmor[2], Efthimios Kaxiras[1,10], Philip Kim[1*]

[1] Department of Physics, Harvard University, Cambridge, MA 02138, USA

[2] Aerospace Engineering and Mechanics, University of Minnesota, Minneapolis, MN 55455, USA

[3] Department of Mathematics, University of Kansas, Lawrence, KS 66045, USA

[4] Department of Materials Science and Engineering, University of Michigan, Ann Arbor, MI 48109, USA

[5] Institute for Quantum Computing and Department of Chemistry, University of Waterloo, Waterloo, ON N2L 3G1, Canada

[6] National Institute for Materials Science, Namiki 1-1, Ibaraki 305-0044, Japan

[7] Department of Physics and Astronomy, Seoul National University, 1 Gwanak-ro, Gwanak-gu, Seoul 08826, Republic of Korea

[8] Department of Materials Science and Engineering, Seoul National University, 1 Gwanak-ro, Gwanak-gu, Seoul 08826, Republic of Korea

[9] School of Mathematics, University of Minnesota, Minneapolis, MN 55455, USA

[10] John A. Paulson School of Engineering and Applied Sciences, Harvard University

Control of the interlayer twist angle in two-dimensional (2D) van der Waals (vdW) heterostructures enables one to engineer a quasiperiodic moiré superlattice of tunable length scale[1-7]. In twisted bilayer graphene (TBG), the simple moiré superlattice band description suggests that the electronic band width can be tuned to be comparable to the vdW interlayer interaction at a 'magic angle'[8], exhibiting strongly correlated behavior. However, the vdW interlayer interaction can also cause significant structural reconstruction at the interface by favoring interlayer commensurability, which competes with the intralayer lattice distortion[9-15]. Here we report the atomic scale reconstruction in TBG and its effect on the electronic structure. We find a gradual transition from incommensurate moiré structure to an array of commensurate domain structures as we decrease the twist angle across the characteristic crossover angle, $\theta_c$ ~1°. In the twist regime smaller than $\theta_c$ where the atomic and electronic reconstruction become significant, a simple moiré band description breaks down. Upon applying a transverse electric field, we observe electronic transport along the network of one-dimensional (1D) topological channels that surround the alternating triangular gapped domains, providing a new pathway to engineer the system with continuous tunability.

In the absence of atomic scale reconstruction, a small rigid rotation of the vdW layers relative to each other results in a moiré pattern, whose long wavelength periodicity is determined by the twist angle. For unreconstructed TBG, atomic registry varies continuously across the moiré period between three distinct types of symmetric stacking configurations: energetically favorable AB and BA Bernal stacking and unfavorable AA stacking (Fig. 1a). This quasiperiodic moiré superlattice, associated with the incommensurability of the twisted layers, modifies the band structure significantly. In the small twist regime, low-energy flat bands appear at a series of magic angles ($\theta_{magic} \leq 1.1°$) where the diverging density of states (DOS) and vanishing Fermi velocity, associated with strong electronic correlation, are predicted[8]. The recent experiment demonstrated the presence of the first magic angle near ~ 1.1° where Mott insulator and unconventional superconductivity were observed[6,7]. The TBG moiré band calculation, however, assumes a rigid rotation of layers ignoring atomic scale reconstruction. Despite the weak nature of vdW interaction and the absence of dangling bonds, recent experimental works on similar material systems suggest there is substantial lattice reconstruction at vdW interfaces, especially at small twist angle close to global commensuration between two adjacent layers[9,10]. Atomic scale reconstruction at vdW interfaces induces significant changes in lattice symmetry and electronic structure[9-15]. In TBG, the interfacial reconstruction is expected to occur by rotating the lattice locally (marked by arrows in Fig. 1b) to form an array of domains with Bernal stacking configurations similar to the schematic diagram shown in Fig. 1b.

Here we employ transmission electron microscopy (TEM) combined with electron transport measurements and first principle calculations to investigate the effect of lattice reconstruction in TBGs with small twist angle $\theta$. We fabricate TBG samples with the twist angle in the range of $0 < \theta < 4°$ by employing the experimental technique reported in previous studies[4-7] (see Supplementary Information (SI) S1 for fabrication details). Figure 1c shows a false-color optical microscopy image of TBG covered with multilayer h-BN on a thin SiN membrane for TEM study. Micromechanical manipulation combined with a microscopic probe allows us to engineer and study TBG with less than 10% variation in length scale (domain size) for more than $10^3$ moiré unit cells with domain size up to ~200 nm.

Formation of the commensurate domains via reconstruction can be probed by mapping the stacking order of TBG using TEM dark field (DF) imaging. This experimental technique employs an aperture to filter a specific Bragg peak $\boldsymbol{g}$ on the diffraction plane, providing a spatial map of the stacking order, as recently demonstrated in bilayer graphene[16]. Figure 1d shows the DF image ($\boldsymbol{g} = 10\bar{1}0$) obtained from a TBG sample with a series of different twist angles. $\boldsymbol{g} = 10\bar{1}0$ DF image obtained from a TBG sample with 0.1° twist shows a tessellation of triangular domains that alternate with mirrored symmetry (AB/BA), matching the periodicity of the moiré pattern. The periodic array of domain structures separated by sharp mirror boundaries indicates an interfacial atomic reconstruction in TBG which is schematically represented in Fig. 1a and b (see SI S2 for details). The domain contrast observed in 0.1° is similar to the previous reports on AB/BA domains formed in chemically vapor deposited bilayer graphene (BLG)[16-19]. As we increase the twist angle, however, we can clearly observe the triangular domain contrast is weakened and becomes close to one-directional fringes (insets of Fig. 1d) which typically appears in the simple rotational moiré structure without reconstruction[20]. This trend suggests that the reconstruction strength, and thus the degree of commensurability within the domains increases with decreasing $\theta$ (see SI S4 and S5).

We study the reconstruction in TBG as a function of $\theta$ by investigating selected area electron diffraction (SAED) in a systematic manner. Figure 1e shows a representative SAED pattern obtained from TBG with $\theta$=0.4°. In this SAED image, we identify two sets of diffraction peaks with 6-fold rotation symmetry. The brighter set corresponds to Bragg peaks from a thicker h-BN layer, and the weaker set of Bragg peaks corresponds to the TBG. A close inspection of the TBG Bragg peaks (right panels of Fig. 1e) reveals that the two main Bragg peaks from the top and bottom graphene layers, marked with red dotted circles, are surrounded by satellite peaks. These satellite peaks are associated with a periodic modulation of much larger scale than that of the atomic lattice[21,22]. Here, the periodicity and strength of the lattice modulation can be quantified by the position and intensity of satellite peaks (SI S4 and S5). SAED patterns obtained from a series of different twist angles (SI S4) show that the configurational arrangement of satellite peaks is not changed as a function of twist angle: all the satellite peak positions in reference to the adjacent Bragg peaks indicated with the wavevectors $\boldsymbol{q_j}$ are equal to the moiré wavevectors defined by the vector difference between the reciprocal lattice vectors $\boldsymbol{g}_{T(B)}$ for the top and bottom layer (the inset of Fig. 1f). This relation indicates that the periodic lattice modulation at the interface is commensurate with the moiré superlattice which is determined by the twist angle $\theta$.

Quantitative analysis of structural reconstruction can be performed by studying the intensity of a satellite peak ($I_{sat}$) and the adjacent Bragg peak ($I_{Bra}$). We note that TBG is an extremely thin system where we can obtain quantitative details of atomic structures via investigating diffraction intensity under the kinematic approximation (see SI S3). Figure 1f shows experimentally measured $I_{sat}/I_{Bra}$ for the first, second and third order Bragg peaks measured in 11 different samples in the range of $0° < \theta < 4°$. For $\theta > 4°$, $I_{sat}$ is under the experimental detection limit, indicating no appreciable reconstruction occurs in this large $\theta$ limit (see SI S4). We find that $I_{sat}/I_{Bra}$ decreases rapidly as $\theta$ increases, following an empirical relation, $I_{sat}/I_{Bra} \sim e^{-\alpha\theta}$ with the constant $\alpha \approx 2.75 \pm 0.3$ (in the unit of inverse degree) obtained from the linear slope presented in Fig. 1f. We note that $\alpha$ describes how fast the reconstruction strength decreases as $\theta$ increases and thus is related to the measure of interlayer interaction of the system, which plays a key role to determine both structural reconstruction and electronic band structure modification.

We study the diffraction peak intensity further to obtain insight to the atomic scale registry and reconstruction as a function of $\theta$. Although the diffraction peak intensity contains quantitative details of atomic reconstruction (see SI S5 for details), achieving complete structural information directly from the diffraction pattern is not trivial due to the lost phase information in the SAED. Here we performed finite element method (FEM) simulation to model the structural reconstruction and compare with the experimental observation. Using the atomic coordinates of the reconstructed

TBG obtained from the FEM model, the SAED pattern can be computed (Fig. S4b) where $I_{sat}/I_{Bra}$ shows close match with experimental data exhibiting similar exponential decaying behavior (open circles in Fig. 1f). We note that the FEM model exhibits a crossover in structural reconstruction across the characteristic angle, $\theta_c \cong 1°$ [15]. For $\theta < \theta_c$, nearly commensurate domains are formed, characterized by atomic registry close to Bernal stacking and separated by sharp boundaries. The width of the domain boundaries are constant in this regime ($\theta < \theta_c$) associated with the formation of solitonic features regardless of the twist angle. For $\theta \gtrsim \theta_c \cong 1°$, the atomic registry in the domains starts deviating from Bernal stacking in large portion making the transition from one domain to the other rather gradual. The crossover in structural reconstruction occurs when the moiré length scale becomes comparable to the domain boundary width[14]. Therefore, the balance between the interlayer energy and intralayer elastic energy forms domains with reduced commensurability, suppressing the formation of solitons.

The electronic structure of the TBG samples can be probed by electrical transport measurement. We measure the electrical conductance $G$ in the TBG devices as a function of the carrier charge density $n$ of the TBG samples tuned by the gate voltage. From the side peak position, we can identify the domain size and the corresponding twist angle (see SI S6). We find that the estimated twist angle is consistent with the targeted twist angle within experimental precision ($\pm 0.3°$). Figure 2a,c (top panels) shows the experimentally measured conductance $G$ of TBG at different temperatures $T$ as a function of $n/n_0$, where $n_0 = 1/A$ and $A$ is the area of the moiré unit cell. Here we show the results obtained from two representative twist angles, $\theta \approx 1.1°$ and $0.5°$, respectively, corresponding to the first and second magic angle predicted in the TBG moiré structure without reconstruction[8]. Near the first magic angle, $G(n=\pm 4n_0)$ exhibits a well-developed zero at low temperatures, corresponding to the single particle gap due to the band hybridization, which is consistent with the previous report[4]. We also observe the correlated Mott gap developing at $n=\pm 2n_0$, corresponding to the half-filled band as shown in the previous study[7]. As $T$ increases, the minimum of $G(n=\pm 4n_0)$ and $G(n=\pm 2n_0)$ exhibits an activating behavior (SI S6) with the activation energy corresponding to 27 meV (18 meV) and 0.30 meV (0.22 meV), for the electron (hole) side, respectively, indicating the corresponding energy gaps are developing at these fillings. Near the second magic angle, however, $G(n)$ exhibit non-zero minima without an activation behavior, at the single particle band filling $n/n_0=\pm 4, \pm 8, \pm 12$, suggesting the absence of the energy gap.

In order to understand the experimentally observed charge transport, we calculate the electronic band structures for the reconstructed TBG structure obtained from the FEM model discussed above. We present the single particle band structure (Fig. 2b) and corresponding DOS (Fig. 2a, middle and bottom panels) for both unreconstructed and reconstructed configurations in TBG with $\theta$=1.1°. In both cases, the DOS near the magic angle sensitively depends on the change of $\theta$. As shown in Fig. 2a (middle and bottom panels), small variation near the magic angle changes the postion and magnitude of DOS peaks, indicating precise tuning of $\theta$ is necessary to optimize the singular behavior of the DOS to promote correlated electronic behavior. We note that the reconstructed TBG exhibits sizable gaps of 32 meV and 26 meV for electrons and holes, respectively, which exhibit better agreement with experimentally observed gaps than those from the unreconstructed TBG where no or much smaller gap are shown (Fig. 2b). We also note that the overall DOS is reduced in the reconstructed band compared to unreconstructed ones as the bands themselves have larger dispersion. Structural reconstruction occurring at the twist regime above or close to the crossover angle ($\theta \geq \theta_c$) seems to change the details of band structure such as single particle gap size and band dispersion. Nevertheless, the essential physics related to the magic angle, i.e., the singular development of the DOS peaks due to the flat band condition, remains qualitatively intact.

At the twist angle regime below the crossover angle ($\theta < \theta_c$), however, the electronic band structure of reconstructed TBG differs significantly from that of unreconstructed TBG. To study the reconstruction effect below the crossover angle regime, we choose the representative twist angle close to the theoretically predicted second magic angle, $\theta \approx 0.5°$. As shown in Fig. 2c (middle), the DOS of the unreconstructed TBG varies extremely sensitively with $\theta$. In contrast, the band structure of the reconstructed TBG in this small twist regime is rather insensitive to variations in $\theta$ (Fig. 2c, bottom), presumably because the atomic registry across the whole sample becomes constant except for the narrow domain boundary regions due to the atomic reconstruction. Moreover, the reconstructed TBG bands exhibit larger dispersion than unreconstructed ones. We note that the overlap between each band becomes minimized, creating Dirac points symmetrically at $n/n_0$=0, ±4, and ±8 in the supercell bandstructure where the DOS is also minimized. This band structure for the reconstructed TBG at $\theta \approx 0.5°$ agrees well with experimental observation. $G(n)$ measured for TBG with $\theta$ = 0.47° at low temperatures exhibits dips at $n/n_0$=0, ±4 and ±8, consistent with the multiple Dirac points formed in the superlattice band as a result of the reconstruction.

More than an order of magnitude suppression in DOS peaks in this small $\theta$ limit compared to those in the first magic angle near 1.1° suggests that no signicant correlated electronic behaviors occur in this regime.

Although the correlated electronic behavior seems to be suppressed in the small twist angle regime ($\theta<\theta_c$), the enhanced atomic and electronic reconstruction creates a network of chiral 1D propagation channels. Under a transverse electric field, the nearly commensurate AB and BA domains are gapped out, leaving topologically protected 1D conduction channels along their boundaries[23-26]. Figure 3a-c show longitudinal resistance $R_{xx}$ as a function of top ($V_{top}$) and bottom ($V_{bot}$) gate voltages for three BLG samples: (i) Bernal stacked BLG ($\theta = 0°$), (ii) a large-angle TBG ($\theta \cong 2.5° > \theta_c$), and (iii) small-angle TBG ($\theta = 0.47° < \theta_c$). For a fixed value of $V_{top}$, $R_{xx}$ ($V_{bot}$) displays a maximum value at the charge neutrality point (CNP). The location of these maxima appears as a diagonal line in the $V_{top}$ - $V_{bot}$ plane, where the average transverse electric displacement field $D = \frac{C_{top}V_{top}-C_{bot}V_{bot}}{2} - D_0$ can be tuned. Here, $C_{top}$ ($C_{bot}$) is top (bottom) gate capacitance and $D_0$ is a small residual displacement field.

Figure 3d shows the displacement-tunable resistance $R_{xx}$ along the CNP line, $R(D)$, normalized by the resistance at the global CNP ($D=0$), $R_0$. For $\theta = 0$, i.e., a Bernal stacked BLG, $R/R_0$ increases rapidly as $|D|$ increases due to the gap opening at the CNP by breaking inversion symmetry[27-29]. For large-angle TBG ($\theta \cong 2.5°$), however, $R/R_0$ decreases as $|D|$ increases. The electronic band of this large-angle TBG can be described by two Dirac cones displaced in reciprocal space with negligibly small interlayer coupling (inset of Fig. 3b)[30,31]. Thus the weakly coupled layers are doped with an equal amount but opposite sign of carriers as $|D|$ increases, making both layers less resistive. In contrast, the electron transport behaviors observed in the small-angle TBG ($\theta = 0.47°$) (iii) are distinctly different from the Bernal stacked BLG (i) and large-angle TBG (ii) discussed above. The effect of the domain structures on transport properties becomes evident when the transverse displacement field is applied to form two topologically-distinct insulating states in AB and BA domains, creating a network of 1D topological channels along the domain boundaries (inset of Fig. 3c). Here, each 1D conducting channel, represented as a simple resistor in the schematic diagram, carries quantum resistance $R_q=h/4e^2$. Experimentally, $R(D)$ measured in the small-angle TBG (iii) exhibits no significant change as a function of $|D|$. The value of $R(D) \approx R_0 \approx 3$ k$\Omega$ is of similar order of magnitude to $R_q$, as expected from a simple network of $R_q$ assuming incoherent mixing of currents at vertices. Similar behaviors were obtained for several small-angle TBG ($\theta < \theta_c$) (see SI S6 for more data).

We also investigate magnetotransport in the small angle TBG ($\theta$=0.47°) with an applied magnetic field $B$ perpendicular to the plane. Figure 3e shows $R_{xx}$ as a function of $B$ and $n$ at fixed $D$. We observe well developed Landau fans originated from the main Dirac peaks at the CNP as well as other Landau fans from the side peaks, indicating development of the fractal spectrum known as Hofstadter's butterfly[2,3,32]. Multiple quantum Hall features fan out from two sets of side peaks, creating periodic quantum hall features when the magnetic flux per supercell $\phi$ becomes integer multiples of the magnetic flux quantum $\phi_0 = h/e$. The most salient feature in these fan diagrams is the horizontal features appearing in all three different displacement field conditions ($D/\varepsilon_0 \sim$ -0.7, 0, and 0.7 V/nm) at $B$ = 5.4 T (marked with black arrows in Fig. 3e). This is the magnetic field corresponding to $\frac{\phi}{\phi_0} = 1$, where the magnetic flux $\phi = BA$ with the area of the moiré unit cell yields $\theta$=0.47°. This estimated value of $\theta$ from the fractal spectrum is consistent with the estimation from the side peak positions (see SI S6). Interestingly, we notice that the observed Hofstadter features becomes substantially weaker as |$D$| increases (see Fig. 3f). This suppression of the quantum Hall features at high displacement field suggests the decrease of Landau gaps in the Hofstadter's fractal energy spectrum. We thus attribute this change of fan diagram to the transition of electronic transport from 2D to 1D network mode with increasing |$D$|, confining the carriers into the topologically protected boundaries of gapped AB/BA domain in structurally reconstructed TBG.

**Acknowledgements** We thank Yuan Cao and Pablo Jarillo-Herrero for important discussions. The authors acknowledge the support of the Army Research Office (W911NF-14-1-0247) under the MURI program. A part of the TEM analysis was supported by Global Research Laboratory Program (2015K1A1A2033332) through the National Research Foundation of Korea (NRF). P.K. acknowledges partial support from the Gordon and Betty Moore Foundation's EPiQS Initiative through Grant GBMF4543 and Llyod Foundation. R.E. acknowledges support from the National Science Foundation Graduate Research Fellowship under Grant No. DGE1745303. K.W. and T.T. acknowledge support from the Elemental Strategy Initiative conducted by the MEXT, Japan and the CREST (JPMJCR15F3), JST. Nanofabrication was performed at the Center for Nanoscale Systems at Harvard, supported in part by an NSF NNIN award ECS-00335765.

**Author Contributions** H.Y. and P.K. conceived the experiments. H.Y. and R.E. performed the experiments and analyzed the data. S.C., S.F. and E.K. performed DFT calculation. K.Z. and E.B.T. conceived and performed the theoretical and FEM analyses. P.C. and M.L. performed mathematical modeling analysis. S.S., R.H., A.W.T., G-C.Y., and M.K. performed TEM data analysis. K.W. and T.T. provided bulk hBN crystals. H.Y., R.E., and P.K. wrote the manuscript. All authors contributed to the overall scientific interpretation and edited the manuscript.

**Author Information** The authors declare no competing financial interests. Correspondence and requests for materials should be addressed to P.K. (e-mail: pkim@physics.harvard.edu).

**Figure legends**

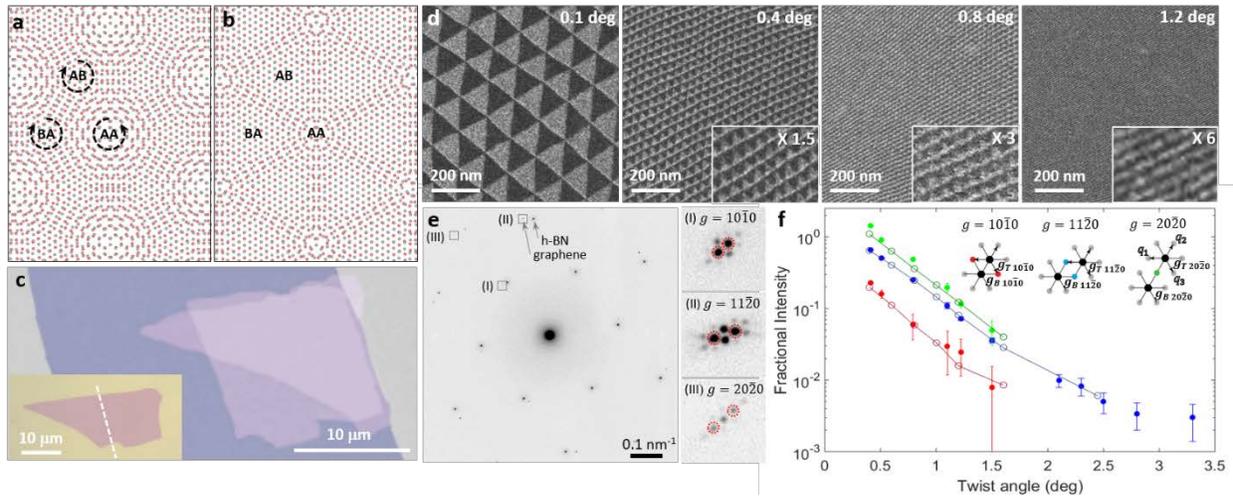

**Figure 1. Atomic scale reconstruction in twisted bilayer graphene (TBG) with controlled twist angle. a, b,** Schematic diagram of TBG before (**a**) and after (**b**) the atomic reconstruction. Periodic rotational modulation of the lattice is represented with the dashed arrows in **a**. **c,** False-color optical microscope image of artificial bilayer graphene with controlled twist angle covered with an h-BN layer. Purple and blue colored regions correspond to graphene and h-BN, respectively. The inset shows optical microscope image of original graphene layer employed to fabricate the twisted bilayer graphene in the main panel. The graphene was torn along the dashed line indicated in the inset. **d**, TEM DF images obtained by selecting the graphene diffraction peak ($g = 10\bar{1}0$) in a TBG with a series of controlled twist angle. The alternating contrast of AB/BA domains is associated with the antisymmetric shift of lattice period in AB and BA domains. **e**. SAED pattern of TBG covered with an h-BN layer (< 5nm) with a twist angle of 0.4°. Diffraction peaks originated from graphene and h-BN are marked with arrows. Each of the graphene peaks marked with black squares and roman numerals are magnified and represented on the rightside to show details of the Bragg and satellite peaks. Red dashed circles indicate the Bragg peaks with the Miller indices shown in the images. **f**. Plot of fractional intensity ($I_{sat}/I_{Bra}$) as a function of twist angle of the TBG. Closed circles with error bars are obtained from experimental SAED patterns, and open circles and line fits are obtained from simulated SAED patterns[15]. Red, blue, and green plots correspond to the satellite peaks marked with red, blue and green closed circles schematically represented in the inset. Bragg peaks of graphene ($g_{T(B)\,10\bar{1}0}$, $g_{T(B)\,11\bar{2}0}$, $g_{T(B)\,20\bar{2}0}$) are represented as black closed circles, and satellite peaks are represented as red, blue, green, and grey circles. Subscripts T and B indicate top and bottom layers, respectively. Three basis vectors ($q_1$, $q_2$, $q_3$) for the lattice modulation index satellite peaks.

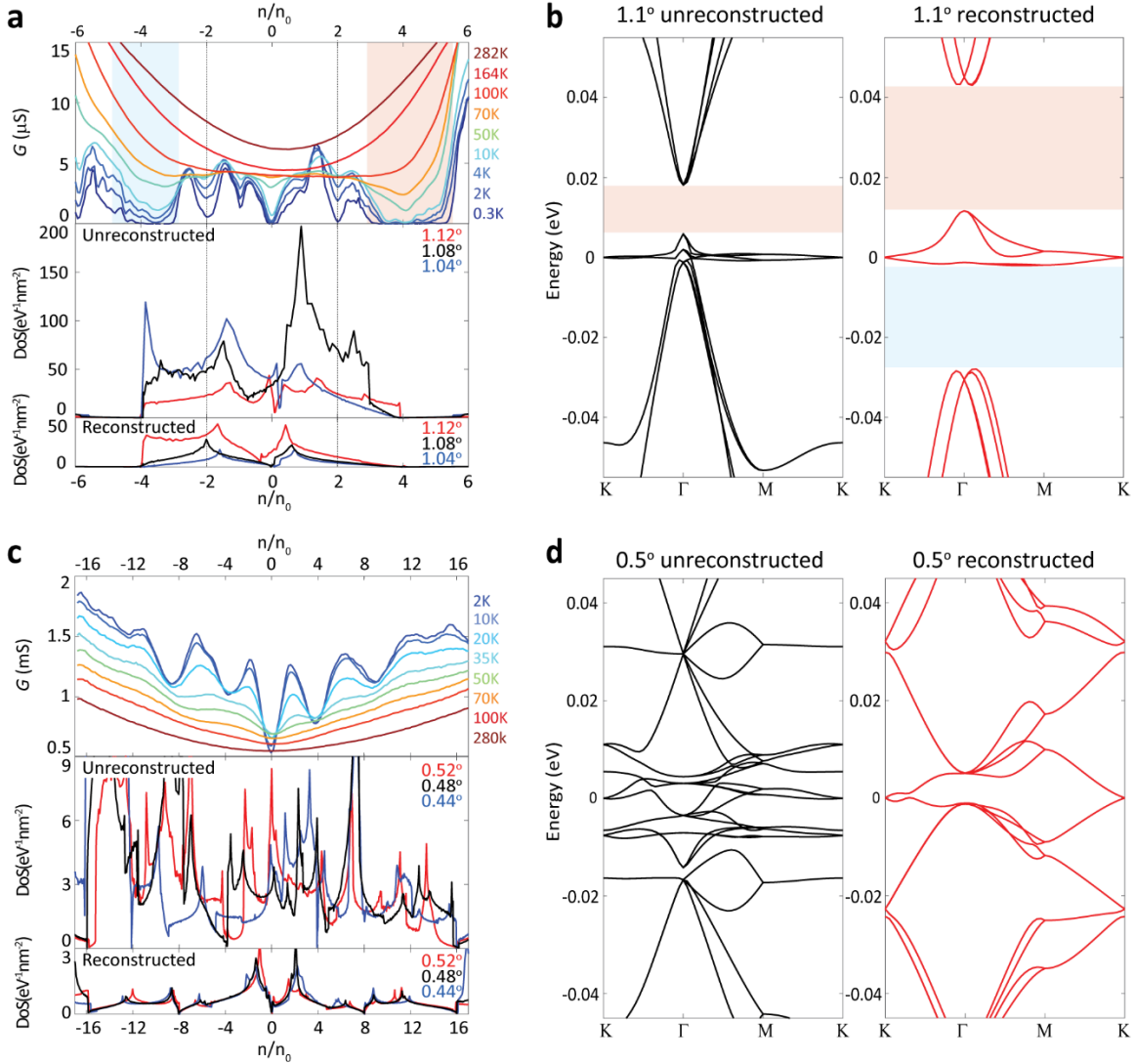

**Figure 2. Electronic reconstruction in twisted bilayer graphene. a**, Temperature dependent conductance $G$ measured from TBG device with $\theta=1.1°$ from 0.3 K to 282K (top). Calculated DoS without (middle) and with (bottom) reconstruction for three angles near 1.1° in units of density normalized by one electron per moiré supercell ($n_0$). **b**, Band structures for the TBG ($\theta=1.1°$) without (left) and with (right) reconstruction. Reconstruction changes the details of the band structure such as single particle gap size, bandwidth, and overall DoS. However, the essential physics of the correlated behavior stays the same. **c**, Temperature dependent conductance $G$ measured from TBG device with $\theta=0.47°$ from 2 K to 280K (bottom). Calculated DoS without (top) and with (middle) reconstruction for three angles near 0.5°. **d**, Band structures for TBG ($\theta=0.5°$) without (left) and with (right) reconstruction. Here reconstruction changes the band structures significantly and explains the single-particle features of gate-dependent conductance.

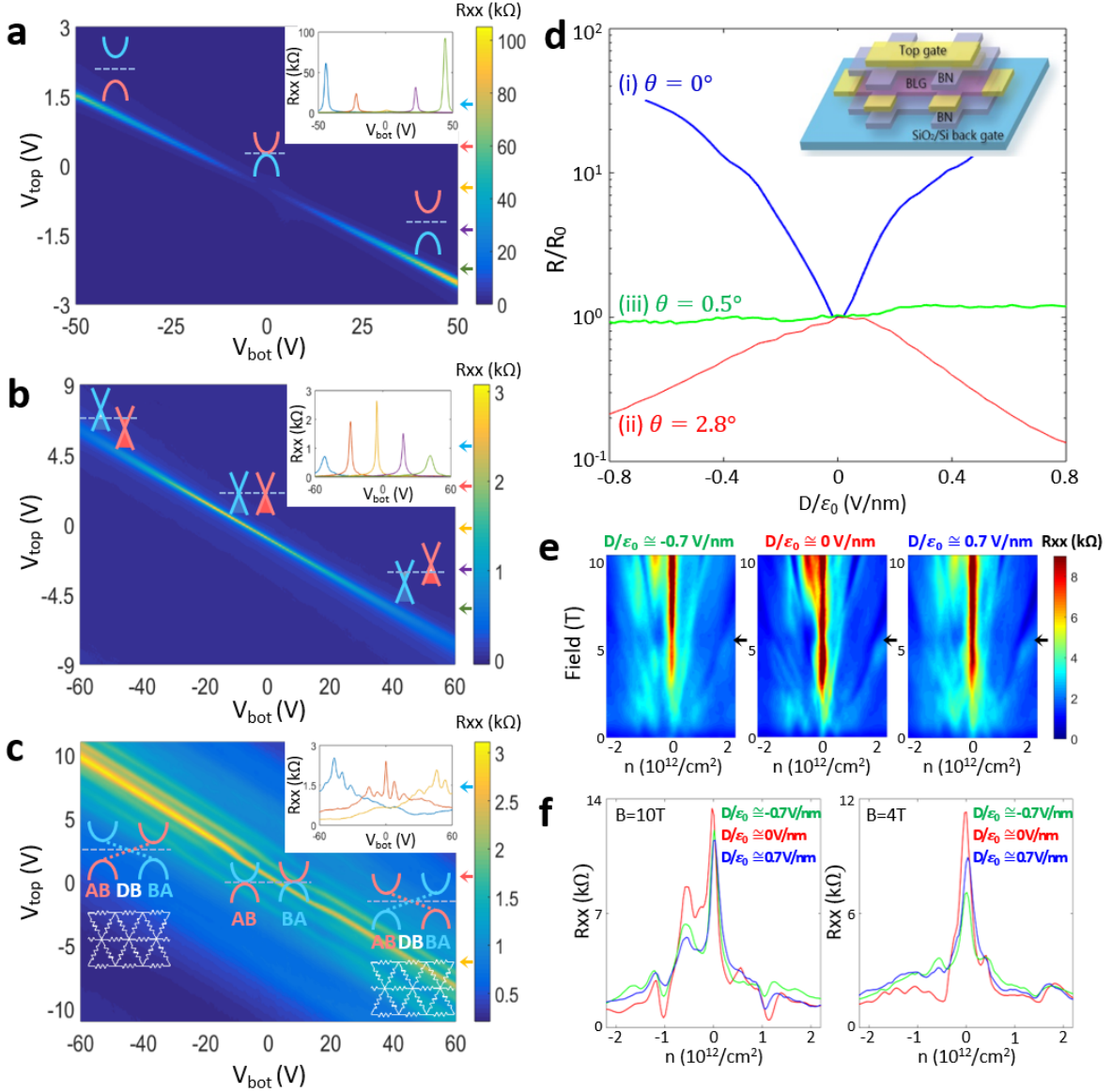

**Figure 3. Transport properties of bilayer graphene with controlled twist angle. a**, **b**, **c**, The top and bottom gate dependence of the longitudinal resistance $R_{xx}$ in Bernal stacked bilayer graphene (**a**), large-angle twisted bilayer graphene (2.8°) (**b**), and small-angle twisted bilayer graphene (0.47°) (**c**). The insets in the top right corner show several line cuts at fixed top gate voltages marked with colored arrows on the right side of the main panel. Schematic band structures in the inset shows how the band structures change as a function of the perpendicular electric displacement field, $D$. Schematic diagram of the triangular network of resistors in **c** represents the current paths generated along the domain boundaries obtained by gapping out AB and BA domains with transverse electric field. DB denotes the domain boundaries. **d**, Normalized Dirac peak resistances as a function of transverse displacement field obtained from **a, b, c**. The inset shows a schematic diagram for a TBG device encapsulated with top and bottom BN. **e**, Displacement field-
14

dependent Landau fan diagram showing longitudinal resistance $R_{xx}$. The fan diagrams shown in the left, middle and right panels were obtained at the displacement field $D/\varepsilon_0 \sim$ -0.7, 0, and 0.7 V/nm, respectively. **f**, Displacement field-dependent longitudinal resistance $R_{xx}$ acquired at constant $B$ =10 T (left) and 4 T (right). Green, red, and blue curves correspond to the displacement field $D/\varepsilon_0 \sim$ -0.7, 0, and 0.7 V/nm, respectively.

## Methods

### TEM experiment

Twisted bilayer graphene was fabricated by tearing single layer graphene into two and engagingthem together with a controlled twist angle using a top BN layer (see SI S1 for details). Using straight edges of the flakes, we intentionally misaligned the top BN layer with the graphene to resolve the diffraction pattern of graphene clearly without overlap with that of BN. In order to detect weak satellite peaks in the diffraction pattern clearly and improve the DF image contrast, it is preferable to have the thickness of the top BN layer and underlying SiN membrane as thin as possible. ~5 nm thick BN and less than 10 nm thick amorphous SiN membrane was used in this work. An 80 kV field-emission TEM (Jeol 2010F) equipped with Gatan One View camera was used for SAED, and DF imaging.

### FEM simulation

Simulations were preformed using a novel continuum-atomistic FEM method for 2D layered heterostructures[33]. A subdivision FEM formulation is used, which provides a smooth parameterization with square integrable curvature. Each layer is treated as a separate mesh that interacts with its neighbor through an interlayer energy density that is integrated across the domain using an efficient discrete-continuum scheme[34]. The mechanical response within a graphene layer is described by a nonlinear elastic model, which consists of a Saint Venant-Kirchhoff membrane term and a Helfrich bending term[35,36]. Non-bonded interactions between the layers are described using a Kolmogorov-Crespi (KC) model, which accounts for registry effects at the interface[37]. A stable equilibrium configuration is obtained by minimizing the total energy of the entire assembly using L-BFGS.

### DFT calculation

The electronic structure of twisted bilayer graphene was modeled using previously obtained ab-initio tight binding parameters[38,39]. The band structures were computed with a standard supercell tight binding Hamiltonian while the density of states (DoS) used an ab initio continuum model[40]. A finite elements method[41] was used on a 60 x 60 k-point grid to generate the DoS profiles. Relaxations were obtained with a continuum model[42] based on the parameters reported previously[43].

**Transport measurement**

For electrical transport measurement, encapsulated twisted bilayer graphene devices with dual gates was fabricated on SiO$_2$/Si substrates. Typical thickness of the top and bottom BN was 15 – 40 nm. We used electron beam lithography to define the top gate, Hall bar, and contact patterns. The top gate was made with Cr/Au (5 nm/70 nm) using BN as the gate insulator, and then the flakes were etched to define a Hall bar pattern using a reactive ion etcher with a mixture of CHF$_3$, Ar, and O$_2$. Finally, 1D side contact to the graphene was made with Cr/Pd/Au (2 nm/10 nm/70 nm)[44]. Heavily doped Si substrate with 285 nm thick SiO$_2$ was used as bottom gate. Transport measurement was performed in a four-probe configuration using a lock-in amplifier at 17.7 Hz with a biasing current of 10 – 30 nA. TBG device with $\theta =1.1°$ was measured in a two-probe configuration using a lock-in amplifier at 17.7 Hz with a bias voltage of 200 μV in a $^3$He cryostat.

*Supplementary Information*

# Atomic and electronic reconstruction at van der Waals interface in twisted bilayer graphene

Hyobin Yoo[1], Rebecca Engelke[1], Stephen Carr[1], Shiang Fang[1], Kuan Zhang[2], Paul Cazeaux[3], Suk Hyun Sung[4], Robert Hovden[4], Adam W. Tsen[5], Takashi Taniguchi[6], Kenji Watanabe[6], Gyu-Chul Yi[7], Miyoung Kim[8], Mitchell Luskin[9], Ellad B. Tadmor[2], Efthimios Kaxiras[1,10], Philip Kim[1*]

[1] Department of Physics, Harvard University, Cambridge, MA 02138, USA

[2] Aerospace Engineering and Mechanics, University of Minnesota, Minneapolis, MN 55455, USA

[3] Department of Mathematics, University of Kansas, Lawrence, KS 66045, USA

[4] Department of Materials Science and Engineering, University of Michigan, Ann Arbor, MI 48109, USA

[5] Institute for Quantum Computing and Department of Chemistry, University of Waterloo, Waterloo, ON N2L 3G1, Canada

[6] National Institute for Materials Science, Namiki 1-1, Ibaraki 305-0044, Japan

[7] Department of Physics and Astronomy, Seoul National University, 1 Gwanak-ro, Gwanak-gu, Seoul 08826, Republic of Korea

[8] Department of Materials Science and Engineering, Seoul National University, 1 Gwanak-ro, Gwanak-gu, Seoul 08826, Republic of Korea

[9] School of Mathematics, University of Minnesota, Minneapolis, MN 55455, USA

[10] John A. Paulson School of Engineering and Applied Sciences, Harvard University

## S1. Sample preparation

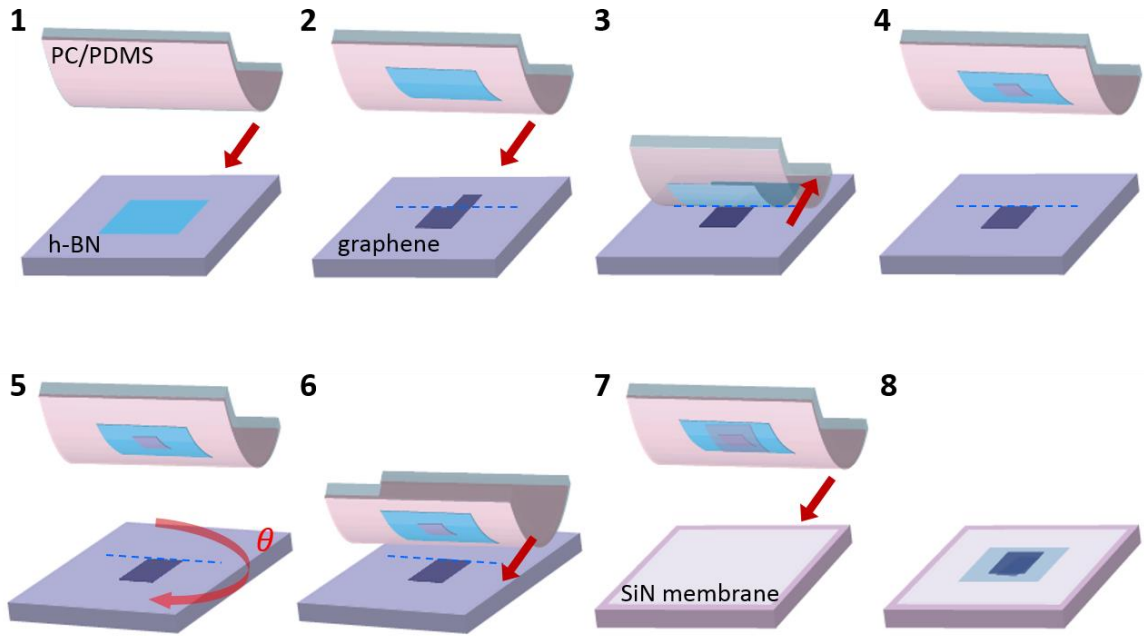

**Figure S1. Schematic diagram of stacking process to fabricate twisted bilayer graphene.**

We employed a van der Waals assembly process to fabricate artificial bilayer graphene with controlled twist angle. A schematic diagram of the stacking process is represented in Fig. S1. First, graphene and h-BN are mechanically exfoliated onto Si/SiO$_2$ (285 nm) substrates. Thickness of the h-BN layer used for TEM study is typically in the range of 5 nm or less, measured by atomic force microscopy. Monolayer graphene was identified by optical contrast and/or Raman spectroscopy. After preparing graphene and h-BN flakes, the thin single crystallite h-BN layer was first picked up at 70°C using adhesive polymer (poly(Bisphenol A carbonate), PC) coated on the stamp made of transparent elastomer (polydimethylsiloxane, PDMS). Then, the h-BN layer was engaged to half of the graphene flake. By lifting the stamp off the substrates, we pick up only the part of the graphene which was covered by top h-BN layer, leaving the remaining part of the graphene on the substrate. After the detachment, the substrate was rotated with controlled angle. Engaging the graphene/h-BN stack on the adhesive polymer to the other half piece of graphene on the substrate makes the artificial bilayer graphene with controlled twist angle[1,2]. The whole stack was picked up and transferred onto thin SiN membrane for TEM study. The stack can be engaged into another h-BN layer in the case that the twist bilayer graphene needs to be encapsulated between top and bottom h-BN layers.

## S2. Dark field TEM images

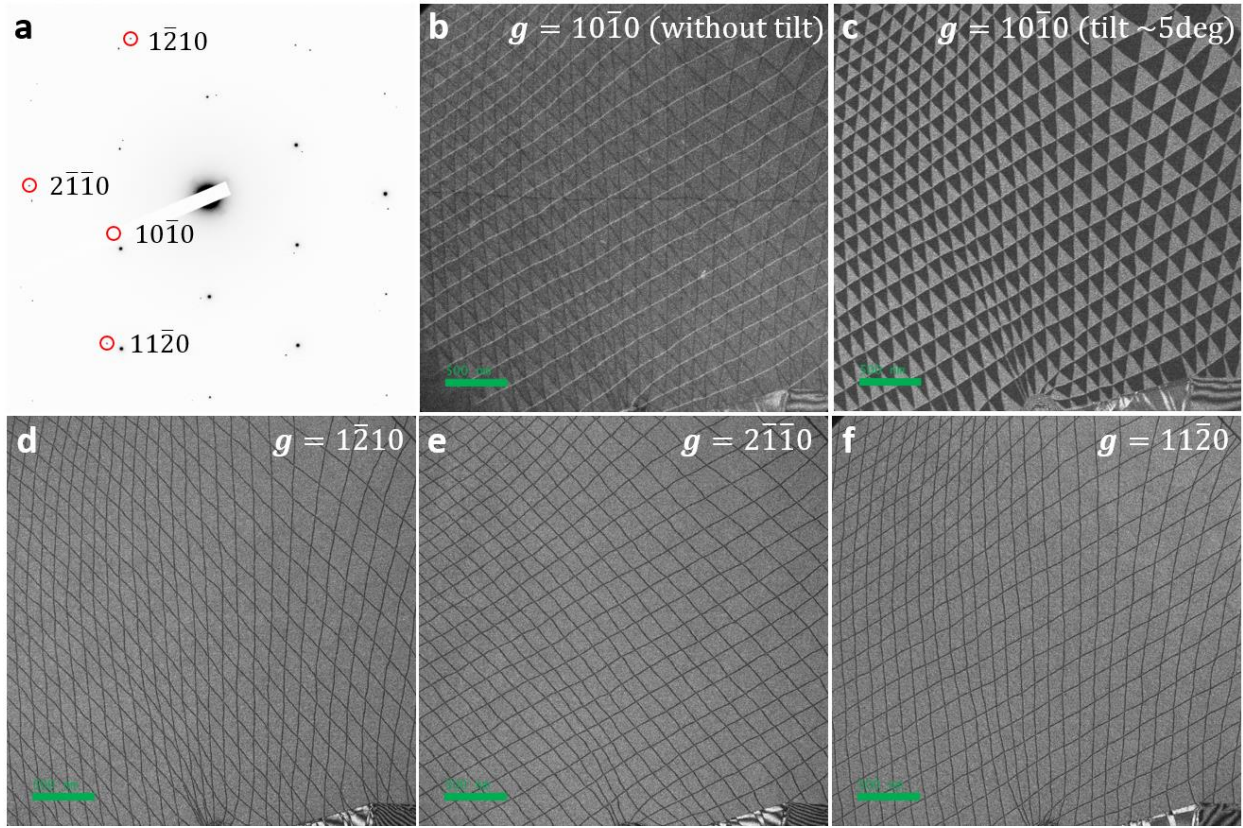

**Figure S2. SAED and DF images of twisted bilayer graphene. a,** SAED obtained from twisted bilayer graphene with the Bragg peaks used to form DF images marked with red circles. **b,** $g=01\bar{1}0$ DF image obtained without tilting the specimen. **c,** $g=10\bar{1}0$ DF image obtained with tilting angle of 5° to enhance the domain contrast. **d, e, f,** $g=1\bar{2}10, 2\bar{1}\bar{1}0, 11\bar{2}0$ DF images obtained from the same region.

We investigate the domain and domain boundary contrast shown in TEM dark field (DF) images. Figure S2a shows the SAED pattern obtained from a TBG sample with a twist angle of 0.1°. The graphene Bragg peaks that were used to obtain DF images (Fig. S2b-f) are marked with red circles and their corresponding indices. Figure S2b shows the $g = 10\bar{1}0$ DF image obtained with the specimen oriented on zone axis (electron beam is incident along the [0001] axis of TBG). AB and BA domains exhibit similar intensity while the domain boundaries are visualized as dark or bright lines depending on the directions of the displacement fields generated along the boundaries, which will be discussed in more detail later in this section. Due to the antisymmetric shift of lattice period in AB and BA domains, AB and BA domains exhibit strong contrast in intensity upon tilting the specimens with respect to the electron beam incident direction as shown in Fig. S2c. Similar type of domain contrast has been reported in bilayer graphene grown by chemical vapor deposition (CVD)[3-6].

Systematic examination of the DF images obtained from a series of different Bragg peaks allows us to study the local displacement induced along the domain boundaries. We note that the domain boundaries can be understood as dislocations with Burgers vectors $\boldsymbol{b}$ which characterize the displacement directions and magnitudes[5]. DF images obtained with three different sets of 2$^{nd}$ order Bragg peaks (Fig. S2d-f) show that the dislocations can be visible ($\boldsymbol{g} \cdot \boldsymbol{b} \neq 0$) or invisible ($\boldsymbol{g} \cdot \boldsymbol{b} = 0$) depending on the angle between Burgers vector $\boldsymbol{b}$ of the dislocations and reciprocal lattice vector $\boldsymbol{g}$ chosen to take the DF image. Each set of dislocations that become invisible in $\boldsymbol{g}$=$1\bar{2}10, 2\bar{1}\bar{1}0, 11\bar{2}0$ DF images have the corresponding Burgers vectors of $\boldsymbol{b_1} = \frac{1}{3}[10\bar{1}0]$, $\boldsymbol{b_2} = \frac{1}{3}[0\bar{1}10]$, $\boldsymbol{b_3} = \frac{1}{3}[\bar{1}100]$, respectively. Burgers vectors of type $\boldsymbol{b} = \frac{1}{3}<10\bar{1}0>$ are partial dislocations, which is consistent with the change from AB to BA domains across each dislocation. Moreover, the Burgers vectors determined above are corroborated by the DF image obtained from the 1$^{st}$ order Bragg peak on zone axis (Fig. S2b) where the dislocations with Burgers vector parallel with $\boldsymbol{g}$ ($|\boldsymbol{g} \cdot \boldsymbol{b}| = \frac{2}{3}$) exhibits brighter intensity while the other dislocations with Burgers vectors that have angle with $\boldsymbol{g}$ of 60° or 120° ($|\boldsymbol{g} \cdot \boldsymbol{b}| = \frac{1}{3}$) show darker intensity. Although the domain and domain boundary contrast described here is similar with the previous report on CVD grown bilayer graphene, the type of displacement induced along the dislocations shows different behavior in each type of sample[3,5,6].

The origin of atomic motion to form commensurate domains and domain boundaries can be inferred by studying the type of displacement induced along the dislocations. In our specimen, Burgers vectors $\boldsymbol{b}$ are mostly parallel with the dislocation line vectors $\boldsymbol{l}$, indicating they carry shear-type displacement. Counter rotation of the top and bottom lattice to form AB and BA domains leaves the domain boundaries carrying the shear-type displacement. On the other hand, previous reports on CVD grown bilayer graphene showed various types of displacement along the domain boundaries presumably depending on their sample preparation method[3-6]. Strip-like domain structures formed on bilayer graphene grown on SiC substrates mostly showed mixed components of shear- and tensile-type displacements[5]. While most of the domain structures formed on bilayer graphene grown on Cu foil did not show a certain tendency in the type of displacement,[3,4,6] the bilayer graphene grown with two additional supporting graphene layers showed alternating triangular domains of AB and BA with the domain boundaries carrying shear type displacement[3], which was similar with our observation.

## S3. Dynamical diffraction simulation

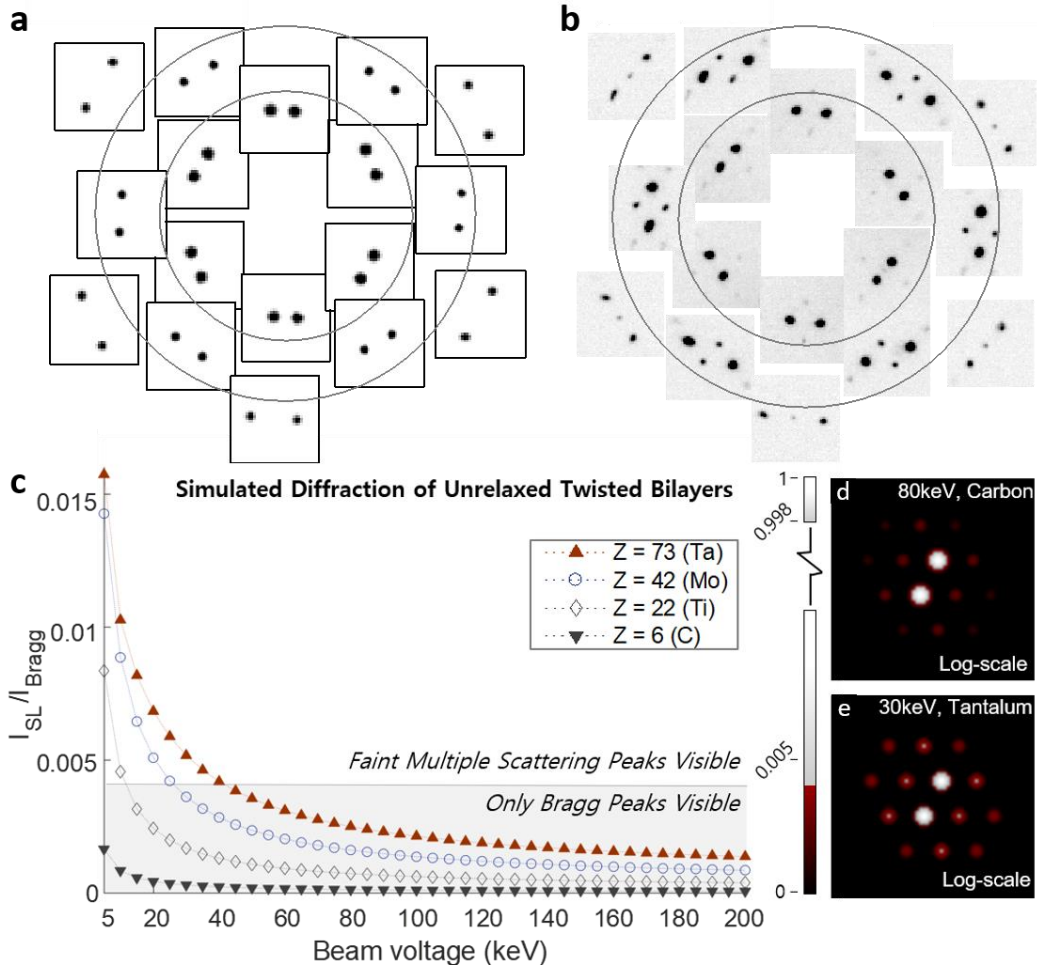

**Figure S3. Dynamical diffraction simulation. a, b,** Comparison of the simulated electron diffraction of twisted bilayer graphene (TBG) without reconstruction (**a**) and experimental electron diffraction of reconstructed TBG (**b**). In the diffraction pattern obtained via multislice simulation from TBG without reconstruction, the satellite peaks are not apparent even without any detector noise or a low-intensity thermal diffuse background. Experimental electron diffraction obtained from the reconstructed TBG clearly shows the satellite peaks with intensities brighter than the noise background, indicating the satellite peaks are not caused by multiple diffraction. Both TBG have the same twist angle (1.1°). **c,** Fractional intensities of satellite peak to $g = 10\bar{1}0$ Bragg peak plotted as a function of acceleration voltage in TEM. Intensities of satellite peaks become stronger when the heavier atomic elements such as Ta, Mo, and Ti substitute the carbon atoms. The cutoff for detectable satellite peak intensities is chosen to be the maximum of the airy function's tail (0.4% of the maximum). **d**, **e**, Log-scaled images of $g = 10\bar{1}0$ Bragg peak of TBG without reconstruction obtained at 80 keV and toy-model structure in which we replace carbon atoms with Ta atoms at 30 keV. The pixels with intensity less than the cutoff level are colored red and the pixels with intensity higher than the cutoff level are colored white.

In TBG, the presence of satellite peaks is a signature of periodic lattice modulations. The atomically thin specimen and low electron scattering cross section of carbon exclude the possibility of multiple scattering features in a diffraction pattern that commonly appear in bulk heterostructures. Multislice simulation allows us to investigate the interaction of probing electrons with specimen including multiple scattering effects. Figure S3a shows a simulated diffraction pattern taken at 80 keV from unrelaxed TBG with the twist angle of 1.1°. The structure of the unrelaxed TBG for the electron diffraction simulation was constructed based on rigid lattice model without any reconstruction. The simulated diffraction pattern (Fig. S3a) appears to be equivalent to the superposition of two sets of single-layer graphene diffraction patterns. On the other hand, the experimental diffraction pattern obtained from TBG (Fig. S3b) with the twist angle of 1.1° contains additional satellite peaks that are noticeably brighter than the noise level.

Fully quantum mechanical multislice simulations with phonons were used to quantify diffraction in unrelaxed TBG to understand when diffraction is no longer approximated as a superposition of constituent layers. Multislice simulation shows no noticeable satellite peaks under typical operating voltages (e.g. at > 80 keV peaks are < 0.13% of the Bragg peak intensity). Heavy atom scattering was also tested for multiple scattering. Using toy-model structures where heavier atomic potentials—Ti, Mo, and Ta—replace carbon sites in unrelaxed TBG (1.1°), we also confirm that multiple scattering peaks at low-voltages are undetectable even for heavy elements such as Ta and Mo—demarcation for visibility is set at 0.4% of the Bragg peaks intensity (i.e. values greater than the tails of an airy disc). Figure S3c plots the relative intensities of a satellite peak and a Bragg peak across beam voltages for several atomic species. We also note that additional simulations with thick (up to 20 nm) boron nitride layers on top of unrelaxed TBG does not enhance the visibility of the superlattice peaks. Multiple scattering is negligible in TBG at typical beam voltages, especially when compared to the superlattice peaks caused by structural relaxation.

Electron diffraction simulations were computed using E. J. Kirkland's multislice code, where the atomic scattering factors are well characterized by a 12-parameter fit of Gaussians and Lorentzians to relativistic Hartree–Fock calculations.[7-9] 32-bit complex wavefunctions were sampled at 4096×4096 pixels over a specimen size of 1125.5×1299.6 Å$^2$ ensuring the scattered wavefunction is characterized out to at least 0.00089 Å$^{-1}$. Frozen phonons were included at an rms atomic displacement for graphene of ~6.3 pm to characterize thermal diffuse scattering[10,11]. Each simulation was averaged over 32 phonon configurations totaling over 640GB of data.

## S4. Analysis of SAED patterns with FEM simulation

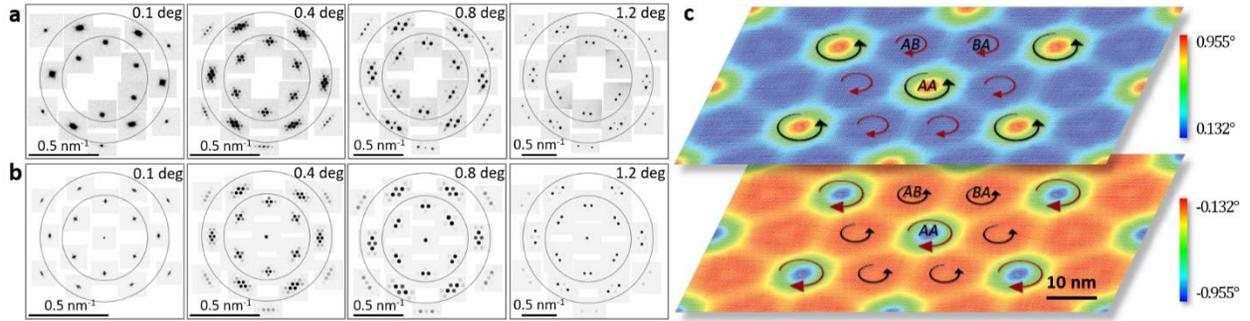

**Figure S4. Twist angle dependence of atomic reconstruction and their resulting structures. a**, Mosaic of TEM SAED patterns obtained from TBG specimens with different twist angles. The small regions where graphene diffraction peaks appear are magnified and placed at their original positions in the diffraction patterns. The scale of diffraction patterns for each angle was adjusted to show all the diffraction peaks clearly (see the scale bars for each panel). Grey circles are drawn to distinguish the three sets of Bragg peaks associated with reflections from $\{10\bar{1}0\}, \{11\bar{2}0\}, \{20\bar{2}0\}$ planes. **b,** Simulated SAED patterns using multiscale FEM presented in the same form as **a**. **c,** Local twist angle map of the top and bottom layers in a TBG with $\theta = 0.6°$ after the reconstruction[12]. Red and black arrows indicate clockwise and counterclockwise rotation of atomic lattices during the reconstruction process to form the domain structures. Within AB and BA domains, the top and bottom layers rotate in opposite directions to decrease the twist angle locally to form a lower energy configuration. Conversely, the reconstruction increases the local twist angle at AA region so that it can move away from the energetically unfavorable AA configuration.

The reconstruction in TBG was studied by investigating SAED. Figure S4a shows the SAED patterns obtained for different twist angle corresponding to the region where we take DF images (Fig. 1d in main text). The satellite peak structures appear near each Bragg peak and can be understood mathematically by decomposing a periodic lattice modulation into a series of Fourier components[13]. In this model, the original and modulated atomic position $\boldsymbol{r}$ and $\boldsymbol{r}'$ can be related by a general expression $\boldsymbol{r}' = \boldsymbol{r} + \sum_j \boldsymbol{\eta}_j \boldsymbol{sin}(\boldsymbol{q}_j \cdot \boldsymbol{r} + \phi_j)$, where $\boldsymbol{\eta}_j$, $\boldsymbol{q}_j$, and $\phi_j$ indicate the modulation amplitude vector, wavevector, and an additional phase term for $j_{th}$ basis vector, respectively[13]. Applying this general model to the TBG, we can define $\boldsymbol{r}_{T(B)}$ and $\boldsymbol{r}'_{T(B)}$ for the top (T) (bottom (B)) layer of graphene. In the diffraction patterns, the Fourier transform of $\boldsymbol{r}'_T + \boldsymbol{r}'_B$ yields a set of satellite peaks around each Bragg peak $\boldsymbol{g}_{T(B)}$. All the satellite peak positions are represented by a linear combination of the modulation wavevectors $\boldsymbol{q}_j$, and schematically shown in the inset of Fig. 1f for the first, second, and third order Bragg peaks (see SI S5 for additional discussion).

The atomic reconstruction in TBG was simulated by a multiscale finite element method (FEM). Figure S4b shows that the diffraction patterns obtained from the simulations for relaxed TBGs are in a good agreement with the experimental data shown in Fig. S4a (quantitative comparison was discussed in main text). Since the FEM model provides a local deformation picture in real space, we find it useful to define a local twist angle $\theta_{loc}^{(T,B)}(r)$ for the top (T) and bottom (B) layers. $\theta_{loc}^{(T,B)}(r)$ is computed from the directions of the C-C bonds of the atomic site in reference to $\theta_{loc}^{(T,B)}(r) = 0$ corresponding to Bernal stacked configuration achieved by untwisting the top and bottom layers symmetrically. For an unrelaxed moiré pattern with a twist angle $\theta_0$, therefore, $\theta_{loc}^{(T)}(r) = -\theta_{loc}^{(B)}(r) = \theta_0/2$ uniformly across the sample. When the TBGs are relaxed, $\theta_{loc}^{(T)}(r) = -\theta_{loc}^{(B)}(r)$ varies over real space with the moiré periodicity. Figure S4c shows $\theta_{loc}^{(T)}(r)$ and $\theta_{loc}^{(B)}(r)$ for $\theta = 0.6°$. A large portion of top and bottom layers rotate locally toward $\theta = 0°$, maximizing the low-energy Bernal stacked area (AB and BA domains). In this configuration, the disregistry between top and bottom lattices concentrates into the domain boundaries. In contrast, in the AA region, the lattices rotate locally up to ~1.9° in the opposite directions compared with that in AB and BA domain region. By increasing the local twist angle in the AA region, the area of the high-energy AA-stacked regions shrink. AB and BA domains are not in perfect registry and can retain a small local twist which we refer as nearly commensurate structure in main text. For the TBG with $\theta = 0.6°$ shown in Fig. S4c, the residual twist angle in Bernal stacked domains $\theta_{loc}^{(T,B)}(r)$ is $\pm 0.13°$, respectively in the top and bottom layers. This residual twist angle quickly vanishes as $\theta$ decreases, while the reverse torque in the AA domain region stays finite. As a result, size of the AA domain and width of the domain boundaries stays constant below the crossover angle ($\theta_c \leq 1°$)[12].

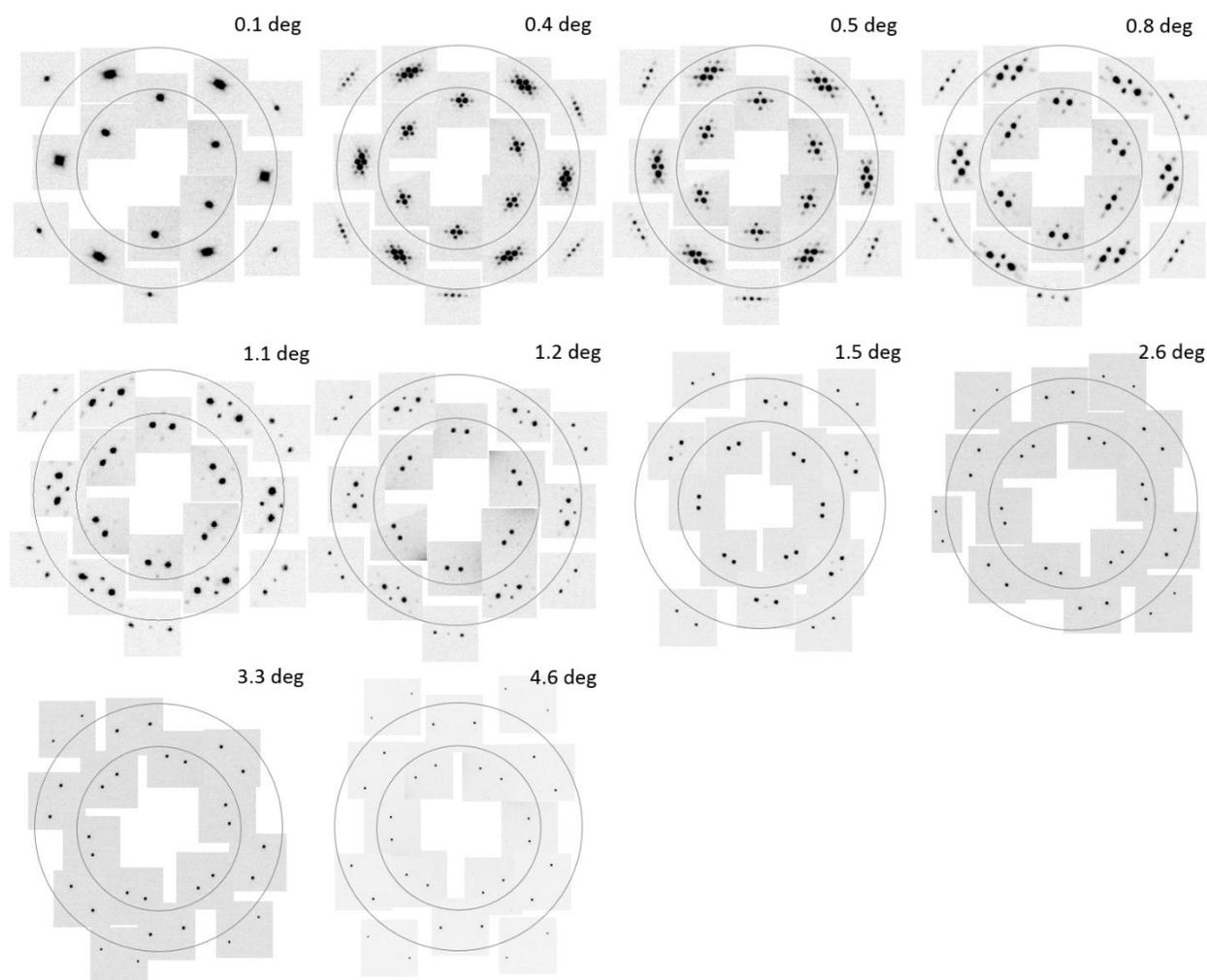

**Figure S5. Mosaic of SAED patterns for a various twist angle from 0.1° to 4.6°.** Scale of each diffraction pattern was adjusted to show both Bragg peaks at each zoomed position.

SAED patterns were obtained from TBGs with a series of different twist angles ranging from 0.1° to 4.6° (Fig. S5). Overall arrangement of the satellite peaks in association with adjacent Bragg peaks stays the same while their intensities decreases as the twist angle increases. The satellite peaks were not observed within the experimental detection limit from the TBG specimen with the twist angle 4.6°.

# S5. Mathematical modeling for reconstructed lattice

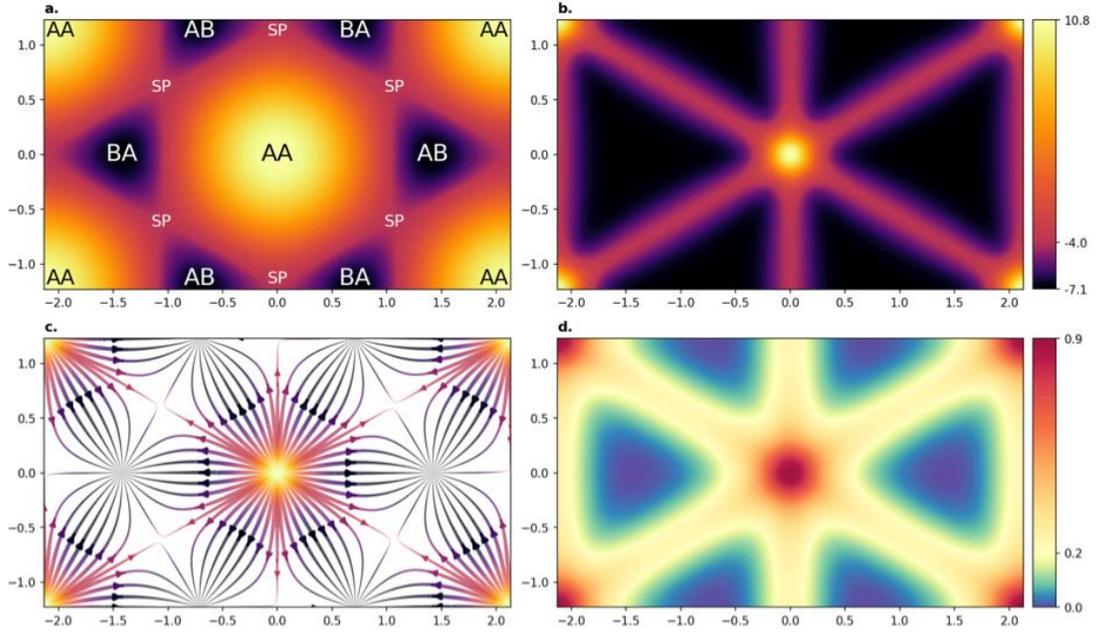

**Figure S6. Registry-dependent lattice reconstruction for a 0.4º twist angle by mathematical modeling. a**, Registry dependence of the generalized stacking-fault energy (GSFE, in mJ/unit cell area) with no relaxation. Energy maxima (AA), minima (AB/BA) and saddle points (SP) are indicated. **b**, Effective GSFE density measured in the simulated relaxed configuration as a function of original registry, showing the growth of nearly commensurate AB and BA domains. **c**. Streamlines of the simulated registry reconstruction vector field. Thickness is proportional to the magnitude of the displacement field. **d**. Local twist angle as a function of original registry.

Due to the incommensurability between the two layers, it is classical that relaxation induces a modulation of the positions of the atoms of layer 1 (resp 2) by a displacement field $u_1(r)$ which has the periodicity of layer 2 (resp 1)[13-15]. The modulated positions of the atoms of the first layer are then given by $r' = r + u_1(r) = r + \sum_j \eta_j \sin(q_j \cdot r + \phi_j)$ where the vectors $q_j$ belong to the reciprocal lattice of the second layer, and vice versa. The displacement $u_1(r)$ is in fact smoothly and uniquely parameterized by the registry of layer 2 relative to layer 1 at position $r$ in the reference (unrelaxed) configuration, which is equal to $\gamma_2 - r$ modulo the lattice of layer 2 where $\gamma_2$ is the registry at the origin.

This observation allows us to formulate a slowly varying interpolant for the displacement field: $\tilde{u}_1(x) = \hat{\eta}(R_{\theta/2}\gamma_2 - (R_{\theta/2} - R_{-\theta/2})x)$. Here $\hat{\eta}$ has the periodicity of the common unrotated lattice structure corresponding to $\theta = 0°$ (Bernal stack) and $R_{\pm\theta/2}$ are the rotation matrices associated with the twist. Note that $u_1(r)$ and $\tilde{u}_1(r)$ coincide on the lattice sites of layer 1 although $u_1(r)$ oscillates at the atomic scale and $\tilde{u}_1(x)$ varies on the scale of the moiré pattern. By further symmetry considerations the modulation of layer 2 can also be deduced from $\hat{\eta}$.

A compact mathematical model for the relaxation of slightly twisted, fully incommensurate bilayers follows from these observations. The interpolant $\tilde{u}_1(x)$ can be used to evaluate the strain of the layer and thus build a continuum description of the elastic response of the layers. The vdW stacking configuration energy is further known to be accurately represented by the local registry-dependent generalized stacking-fault energy (GSFE)[16]. This leads to a complete description of the relaxation mechanism, regardless of the commensurability of the system, as the minimization of the total energy per unit area of the system with respect to the smooth periodic field $\hat{\eta}$[17]. Figure S6 shows the result of this energy minimization procedure driving the system towards commensuration from this registry-based point of view.

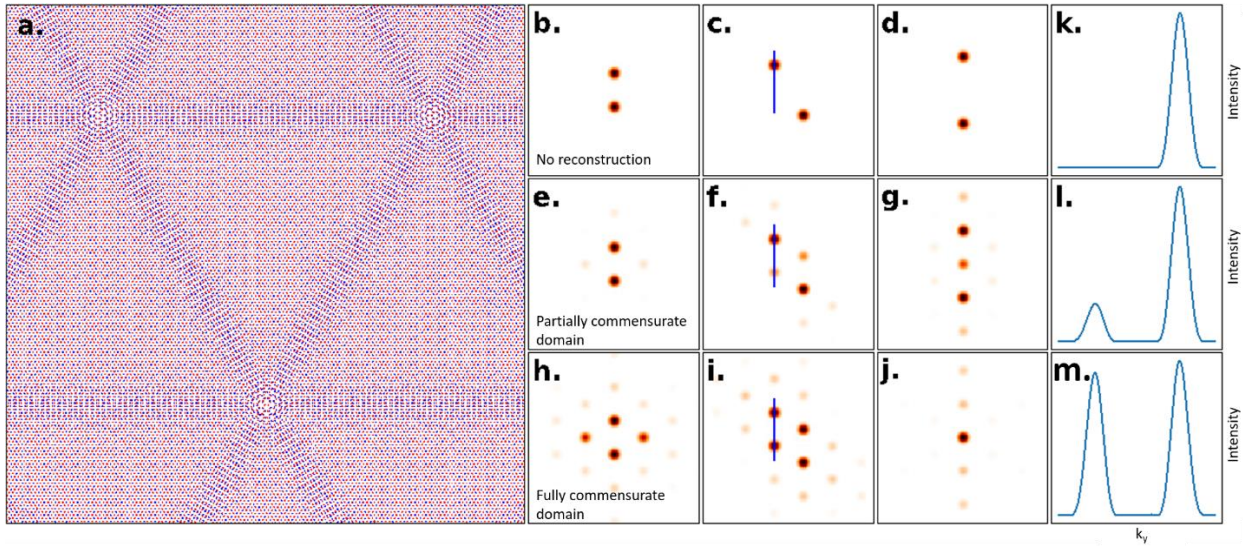

**Figure S7. Reconstructed lattice structures and their simulated diffraction patterns for a 1.1º twist angle with tunable commensurability by mathematical modeling**. **a**, modeled lattice structure with fully commensurate domains (artificial 20-fold increase of the GSFE). **b, c, d**, Diffraction peaks appearing around g=$10\bar{1}0$, $11\bar{2}0$, $20\bar{2}0$ Bragg peaks without any reconstruction. **e, f, g**, Diffraction peaks appearing around g= $10\bar{1}0$, $11\bar{2}0$, $20\bar{2}0$ Bragg peaks from the reconstructed lattice with partially commensurate domains. **h, i, j**, Diffraction peaks appearing around g=$10\bar{1}0$, $11\bar{2}0$, $20\bar{2}0$ Bragg peaks from the reconstructed lattice with fully commensurate domains **k, l, m**, Line cut of the diffraction peak intensity along the line represented in **c, f, i**.

By tuning artificially the GSFE, we can simulate systems with varying degree of commensuration within the domains, allowing us to study the effect on the diffraction pattern by direct kinematic simulations as shown in Fig. S7. We observe that the intensity of the satellite peaks is highly sensitive to the relaxed structure with strong differences between unrelaxed, partially, and fully commensurate domains. Therefore, the close match between the FEM modeling and experimental diffraction patterns strongly suggest that the model describes the atomic scale relaxation of the vdW lattices reasonably well.

## S6. Transport

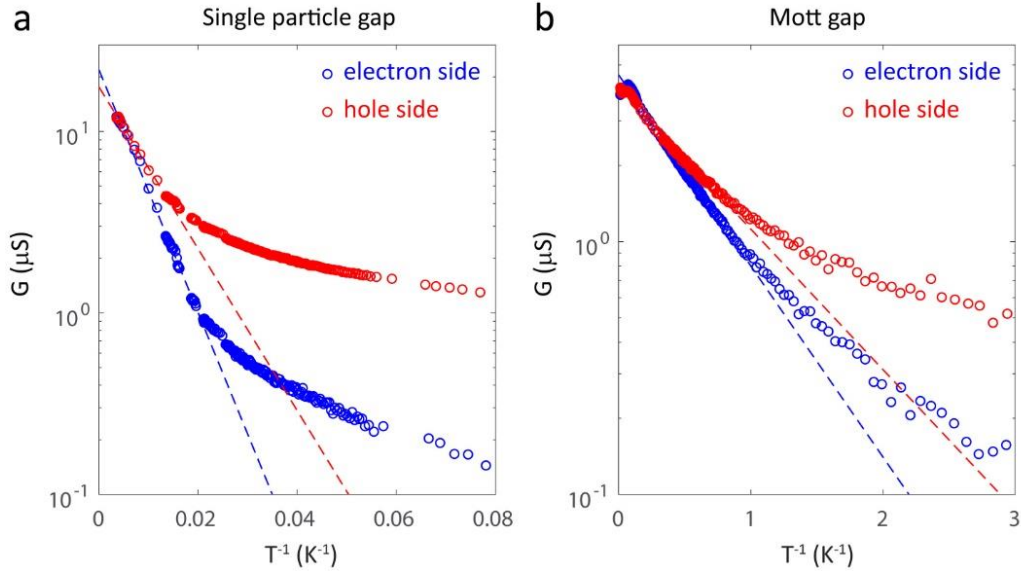

**Figure S8. Measurement of thermal activation gap. a**, **b**, Temperature dependence of conductance minimum for the single particle gap (**a**) and Mott gap (**b**) states. Blue and red open circles denote the experimentally measured values on electron and hole sides, respectively. Dashed lines are Arrhenius fits to the data.

We study the temperature dependence of conductance minimum in the electron and hole sides for the single particle gaps and Mott gaps (Fig. S8a, b). For the single particle gaps occurring at the carrier density $n=\pm 4n_0$, Arrhenius fits to the data, $G \sim \exp(\frac{-E_g}{2k_B T})$, where $k_B$ is Boltzmann constant, yield the thermal activation gap $E_g$ of 27 meV (18 meV) for the electron (hole) side. The temperature dependence of conductance $G$ for correlated Mott gap states at the carrier density $n=\pm 2n_0$ exhibits thermally activating behavior below 15K and metallic behavior above 15K. Arrhenius fits to the data in the thermally activating regime yield the activation gap $E_g$ of 0.30 meV (0.22 meV) for the electron (hole) side, consistent with previous report[18]. We note that the temperature dependence of conductance $G$ for both single particle and Mott insulating states shows deviation from the Arrhenius fit as it approaches to low temperature regime. The deviation from the activating behavior indicates that the charge transport is limited by the variable range hopping where the temperature dependence follows the relation $G \sim \exp(\frac{T_0}{T})^{\frac{1}{3}}$. Similar behaviors have been reported elsewhere[18].

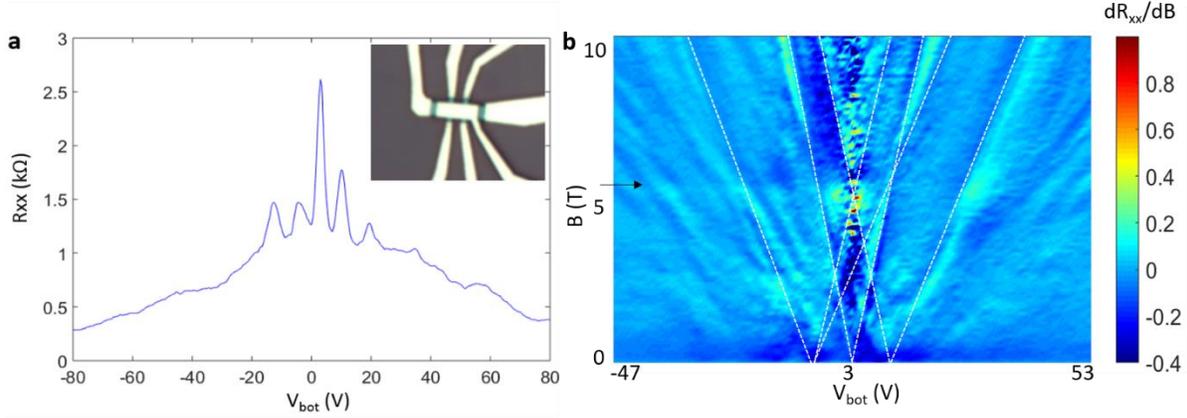

**Figure S9. a**, Plot of $R_{xx}$ as a function of bottom gate voltage with the top gate fixed at 0V without magnetic field. The inset shows optical micrograph of device. **b**, Landau fan diagram represented in the form of derivative of the magnetoresistance $dR_{xx}/dB$. Black arrow indicates the position where the magnetic flux per supercell becomes the magnetic flux quantum $\phi_0 = h/e$.

Figure S9a shows longitudinal resistance $R_{xx}$ of small angle TBG (θ=0.5°) as a function of bottom gate voltage $V_{bot}$ (top gate voltage $V_{top}$ was fixed at 0 V). In addition to the main Dirac peak observed at $V_{bot} \cong 3V$, two sets of side peaks appear on both sides of Dirac peak symmetrically. The 1$^{st}$ set of side peaks appears at a total carrier density $n \approx \pm 5.2 \times 10^{11} cm^{-2}$ which corresponds to the density of 4 electrons per superlattice unit cell assuming 4-fold degeneracy associated with spin and valley. Thus, we can estimate the area of the superlattice unit cell $\frac{\sqrt{3}}{2}\lambda^2 = \frac{4}{n} = 770 \ nm^2$ and the superlattice unit length $\lambda$ of 29.8 nm. From the relation of superlattice unit length ($\lambda$), lattice constant of graphene ($a$), and the twist angle ($\theta$), $\lambda = \frac{a}{2\sin(\theta/2)}$, the twist angle of the specimen can be estimated to be 0.47°. Well-defined side peak features shown in Fig. S9 indicate uniformity of the domain structures within the device. We also estimate the twist angle from Hofstadter butterfly spectrum.

Figure S9b shows Landau fan diagram obtained from the derivative of the magnetoresistance $dR_{xx}/dB$. As discussed in the main text, periodic quantum hall feature appears whenever the magnetic flux per supercell $\phi$ becomes integer multiples of the magnetic flux quantum $\phi_0 = h/e$. $\phi = \phi_0$ occurs at 5.4 T as marked with black arrow in the fan diagram. Using the relation $\phi = \phi_0 = BA = 5.4 \ T \times \frac{\sqrt{3}}{2}\lambda^2$, the superlattice unit length $\lambda$ is estimated to be 29.7 nm which is similar with the value estimated from the side peak positions at B=0 T.

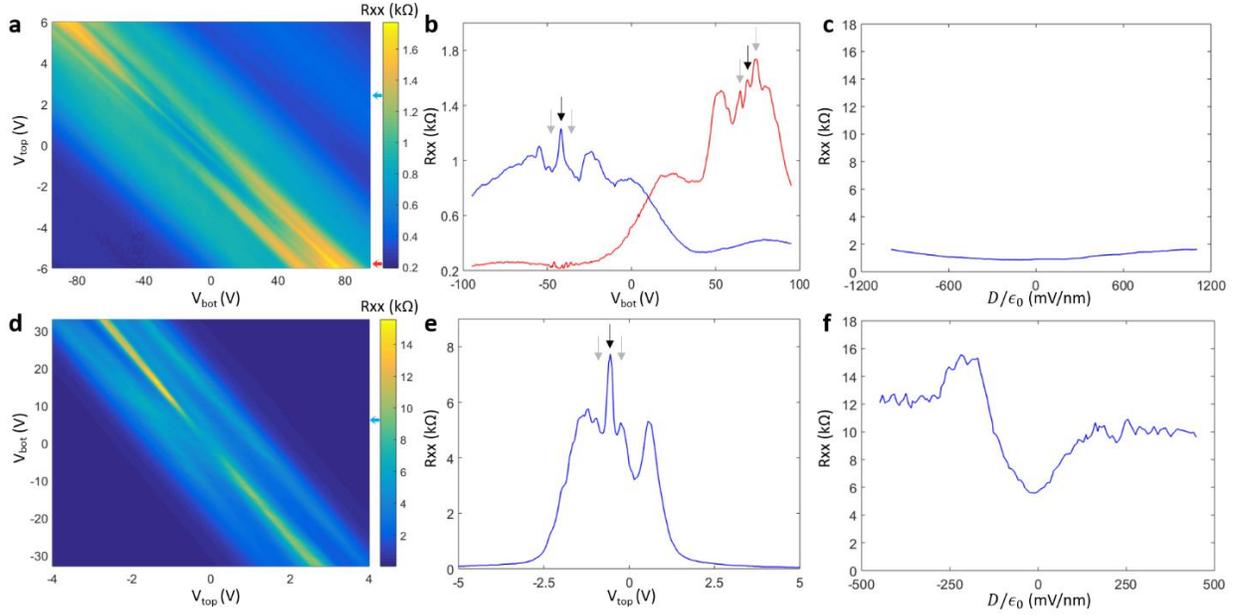

**Figure S10. Dual gate dependence of longitudinal resistance $R_{xx}$. a**, The top and bottom gate dependence of the longitudinal resistance $R_{xx}$ in TBG with a twist angle of 0.41° (specimen S1). **b**, Two line cuts at fixed top gate voltages marked with colored arrows on the right side of the 2-D map shown in **a**. The black and grey arrows indicate the position of main Dirac peak and the 1st set of side peaks, respectively. **c**, Plot of longitudinal resistance $R_{xx}$ at charge neutrality point (CNP) as a function of transverse displacement field obtained from a. **d**, The top and bottom gate dependence of the longitudinal resistance $R_{xx}$ in TBG with a twist angle of 0.37° (specimen S2). **e**, A line cut at a fixed bottom gate voltage marked with blue arrow on the right side of the 2-D map in d. The black and grey arrows indicate the position of main Dirac peak and the 1st set of side peaks, respectively. **f**, Plot of longitudinal resistance $R_{xx}$ at CNP as a function of transverse displacement field obtained from **d**.

Electronic transport along the triangular network of one-dimensional (1D) topological channels was studied from two additional small-angle TBG specimens: specimen S1 (Fig. S10a-c) and specimen S2 (Fig. S10d-f). Figure S10a and d show the top and bottom gate dependence of the longitudinal resistance $R_{xx}$ of the two devices fabricated on S1 and S2. Both devices exhibit a main Dirac peak at the charge neutrality point (CNP) and multiple sets of side peaks. To estimate the twist angle, we plot the line cuts at fixed top/bottom gate voltages as shown in Fig. S10b and e. The line cuts are chosen from the 2-D maps to determine the location of CNP peaks and the 1st sets of side peaks as marked with black and grey arrows (Fig. S10b and e). The twist angles estimated from the position of the 1st side peaks for S1 and S2 were 0.41° and 0.37°, respectively.

In this small twist angle regime, the triangular network of 1D topological channels can be developed by applying a transverse electric field.

Figure S10c and f show the plot of longitudinal resistance $R_{xx}$ at CNP as a function of transverse electric displacement field. S1 shows channel resistance at the global CNP of $R_0 = 0.9$ k$\Omega$ and exhibits slight increase in the resistance up to $R(D) = 1.6$ k$\Omega$ at the maximum displacement field we applied on S1 ($|D| \approx 1000$ mV/nm). S2 shows channel resistance at the global CNP of $R_0 = 6$ k$\Omega$. The channel resistance at CNP increases as higher displacement field is applied ($|D| < 200$ mV/nm) and saturate to constant values at high displacement field ($|D| > 250$ mV/nm) corresponding to $R(D) = 10$ k$\Omega$ and $12$ k$\Omega$ for positive and negative sides, respectively. No significant change in the channel resistance at high displacement field, and the saturating value of the resistance of similar order of magnitude to $R_q = \frac{h}{4e^2} \cong 6.4$ k$\Omega$, suggest electronic transport across the triangular network of 1-D channels.